# Nanodiamonds-induced effects on neuronal firing of mouse hippocampal microcircuits

L. Guarina [1], C. Calorio [1], D. Gavello [1], E. Moreva [2], P. Traina [2], A. Battiato [4], S. Ditalia Tchernij [3,4], J. Forneris [4,3], M. Gai [5], F. Picollo [3,4], P. Olivero [3,4], M. Genovese [2,4], E. Carbone [1], A. Marcantoni [1], V. Carabelli [1,*]

[1] Department of Drug and Science Technology, NIS Center, University of Torino, Corso Raffaello 30, 10125 Torino, Italy

[2] Istituto Nazionale Ricerca Metrologica, Strada delle Cacce 91, 10135 Torino, Italy

[3] Department of Physics and "NIS" inter-departmental centre, University of Torino, Via P. Giuria 1, 10125 Torino, Italy

[4] Istituto Nazionale di Fisica Nucleare, sezione di Torino, Via P. Giuria 1, 10125 Torino, Italy

[5] Dept. of Molecular Biotechnology and Health Sciences, University of Torino, Via Nizza 52, 10126 Torino, Italy

* *Corresponding author:* Valentina Carabelli
  Department of Drug Science and Technology
  Corso Raffaello 30
  10125 - Torino, Italy
  tel.: +39.011.670.8488
  fax: +39.011.670.8498
  e-mail: valentina.carabelli@unito.it

*Running title:* Nanodiamonds effects on hippocampal neurons excitability

**Acknowledgements**

We are grateful to Prof. A. Hernandez-Cruz for critical revision of the manuscript and to Dr. Claudio Franchino for cell preparation.

Ion beam irradiations were performed at the AN2000 accelerator of the Legnaro National Laboratories of the Italian Institute of Nuclear Physics (INFN) within the "Dia.Fab." beamtime, which is gratefully acknowledged.

**Author Contributions statement**

LG, DG, CC contributed to the electrophysiological and confocal trials, data analysis

AB, SDT, JF, FP contributed with the preparation and preliminary characterization of the ND samples

MGai contributed to collect the confocal images and general revision of the manuscript

EM, PT, JF, DT contributed to the experimental work on ND

VC, MG, AM, PO contributed to planning the experiments and wrote the manuscript

EC contributed to planning, analyze and interpret the AP recordings on single neurons. EC also helped with a critical revision for important intellectual content of the manuscript

**Additional Information**

This work has been supported by:

"D-Music project", funded by San Paolo Foundation, Progetti Ateneo, to VC.

"DIESIS" project funded by the Italian National Institute of Nuclear Physics (INFN) - CSN5 within the "Young research grant" scheme to JF.

"DIACELL" project funded by the Italian National Institute of Nuclear Physics (INFN) - CSN5 to FP and PO.

Coordinated Research Project "F11020" funded by the International Atomic Energy Agency (IAEA), and "MiRaDS" project funded by the CRT Foundation, to PO.

MeV He implantations were performed within the "Dia.Fab." experiment at the INFN-LNL laboratories (INFN).

"MiRaDS" project funded by the CRT Foundation to PO.

**Competing Financial Interests statement**

The authors declare no competing financial interests.

**ABSTRACT**

Fluorescent nanodiamonds (FND) are carbon-based nanomaterials that can efficiently incorporate optically active photoluminescent centers such as the nitrogen-vacancy complex, thus making them promising candidates as optical biolabels and drug-delivery agents. FNDs exhibit bright fluorescence without photobleaching combined with high uptake rate and low cytotoxicity. Focusing on FNDs interference with neuronal function, here we examined their effect on cultured hippocampal neurons, monitoring the whole network development as well as the electrophysiological properties of single neurons. We observed that FNDs drastically decreased the frequency of inhibitory (from 1.81 Hz to 0.86 Hz) and excitatory (from 1.61 Hz to 0.68 Hz) miniature postsynaptic currents, and consistently reduced action potential (AP) firing frequency (by 36%), as measured by microelectrode arrays. On the contrary, bursts synchronization was preserved, as well as the amplitude of spontaneous inhibitory and excitatory events. Current-clamp recordings revealed that the ratio of neurons responding with AP trains of high-frequency (fast-spiking) versus neurons responding with trains of low-frequency (slow-spiking) was unaltered, suggesting that FNDs exerted a comparable action on neuronal subpopulations. At the single cell level, rapid onset of the somatic AP ("kink") was drastically reduced in FND-treated neurons, suggesting a reduced contribution of axonal and dendritic components while preserving neuronal excitability.

**INTRODUCTION**

Among the multitude of fairly novel materials for scientific, technological and clinical applications, diamond nanocrystals (or nanodiamonds, NDs) earned a solid reputation and vast interest due to their unique features, such as low cytotoxicity[1], the possibility of stable chemical functionalization and extreme mechanical properties (robustness, low friction coefficient)[2]. The diamond lattice can host a large number of optically active defects[3], the most common and widely employed of which is represented by the negatively charged nitrogen vacancy complex (also referred as $NV^-$ center). This system is characterized by a wide excitation spectrum (500-600) nm, emission in the red range (600-800 nm) with zero phonon line (ZPL) at 638 nm. The neutral charge state of the defect (i.e. the so-called $NV^0$ center, with ZPL emission at 575 nm) is not equally appealing, due to a less convenient electronic structure to implement protocols of local sensing of electromagnetic fields. NDs incorporating $NV^-$ centers provide a stable luminescent label suitable for different types of bio-imaging and bio-sensing applications [4-14]. A significant advantage of fluorescent nanodiamond (FNDs) is related to their photostability, resistance to bleaching or quenching phenomena[15], that allows their monitoring along neuronal branches with high spatio-temporal resolution[13], as well as to perform long-term cell tracking[11,16]. More specifically, the peculiar structure of the spin-dependent radiative transitions of the NV$^-$ centers allows the optical detection of weak electro-magnetic fields and small temperature variations within the biological samples under exams, by means of Optically Detected Magnetic Resonance (ODMR), thus disclosing a range of new perspectives in cell sensing with unprecedented spatial resolution and sensitivity[4,7,9,17,18]. Patterned ND networks and ND coatings also represent excellent substrates for neuronal cultures, preserving intrinsic neuronal excitability as well as spontaneous synaptic transmission, even though nanoparticle size and their surface features may affect neuronal adhesion[19-20]. However, if the bulk material is non-toxic and inert, when dealing with diamond at the nanoscale, quite heterogeneous results emerge. FNDs have been proven to be suitable for long-term *in vivo* imaging in *Caenorhabditis elegans*, without causing toxicity[21-22]; in cancer and stem cells, division and differentiation are not affected by 100 nm NDs[23]. In a different experimental model, airway epithelial cells, functionalization may reduce ND inflammogenicity [24], while a remarkable dose-dependent decrease of neurite length of central and peripheral cultured neurons after FNDs seeding has been demonstrated[25].

Focusing on the interaction of FNDs with hippocampal neurons, in this work we combined patch-clamp and microelectrode array (MEA) recordings to evaluate their functional implications on hippocampal neurons functionality. On mature neurons, FNDs application drastically impairs AP spontaneous firing, as well as the occurrence of spontaneous miniature synaptic currents. On the contrary, the network synchronism, the balance between excitatory and inhibitory synapses and the proportion of fast- versus low-spiking neurons are not affected by FNDs. At the single-cell level, the AP shape analysis suggests that FNDs alter the cell morphology rather than modifying the passive membrane properties and the cell responsiveness to applied electrical stimuli. Hippocampal neurons functionality is also preserved following ODMR test measurements on cell-internalized FNDs.

**RESULTS**

**FND characterization**

The size of FNDs was estimated by means of SEM imaging, after finalizing their preparation. Figs 1a and 1b show typical SEM micrographs, where most of the nano-particles present the expected dimensions, namely below 100 nm. The powders are classified as Ib type, with a nominal substitutional N concentration of 10 ÷ 100 ppm, and contain a low amount of native

NV⁻ centers. Ion beam irradiation (Fig. 1c) and a subsequent thermal annealing were performed to increase the number of NV⁻ centers in each nanocrystal (see Methods). Room temperature photoluminescence (PL) spectroscopy was performed with the purpose of assessing the spectral features of the ND emission. A confocal microscope was employed to this scope, with 532 nm laser excitation and a single-photon-sensitive avalanche Silicon detector. The typical spectrum of NV⁻ centers in NDs is reported in Fig. 1d: consistently with what reported in literature [26-27], the 638 nm zero phonon line and relative phonon bands are observable.

No features involving organic contaminants or graphite are observable, thus confirming the effectiveness of the cleaning procedure (see Methods).

**FND internalization**

FNDs internalization mechanisms and their related motional dynamics inside cells have been previously investigated in various experimental models (HeLa cells, embryo hippocampal neurons, as well as lung cancer cells and embryonic fibroblasts) [22-23,28,29]. Here we assessed FNDs internalization by means of confocal microscopy. Embryo hippocampal neurons were incubated with 40 µg/ml FNDs[25], being this concentration far below the threshold of cytotoxicity (i.e. 250 µg/ml)[11]. After 24 hours, the cytoplasmic labelling dye (CellTracker™ Green CMFDA, ThermoFisher,) was added to the medium. This allowed to identify the cell boundaries together with the internalized FNDs, characterized by a red fluorescent emission. Representative images are shown in Fig. 2a. It is worth noting that since FNDs aggregation may occur in physiological culture media[17], detection of single FNDs may be hindered using our acquisition system.

To exclude the possibility that the FNDs concentration in neurons declines over time, we performed FNDs seeding at two different stages of culture development: at 7 DIV (young cultures) and at 14 DIV (mature cultures) and measured the number of internalized FNDs at 18 DIV, that was respectively after 11 and 4 days of incubation (Fig. 2b). We found that longer periods of incubation (11 vs. 4 days) increase the number of internalized FNDs from $14.6 \pm 3.6$ (4 days) to $29.4 \pm 2.9$ (11 days) ($p<0.01$) (Fig. 2c).

**Firing frequency of the hippocampal network is differently affected by FND administration at early and late stages of development**

Hippocampal neurons, when cultured on microelectrode arrays (MEAs), create a network that exhibit spontaneous firing activity which is characterized by different patterns along with culture maturation: asynchronous firing is typical of younger neurons (7-13 days in vitro,

DIV), while in elder cultures activity becomes highly synchronized and organized into bursts (> 14 DIV)[30]. Here, by means of microelectrode arrays (MEAs), we compared the firing frequency of hippocampal neurons under control conditions and after exposure to FNDs. FNDs seeding was performed at two different stages of culture development, respectively at 7 DIV (young cultures) and at 14 DIV (mature cultures). Unless otherwise specified, firing activity was measured at 18 DIV.

Representative recordings are shown in Fig. 3 respectively for control neurons (a), for neurons incubated with FNDs at 7 DIV (b) and neurons incubated with FNDs at 14 DIV (c). Regardless of the exposure stage of FNDs, we found no changes in the number of spontaneously active neurons, as demonstrated by the unaltered percentage of electrodes exhibiting firing activity: $49 \pm 7$ % in control (n=26 MEAs), $41 \pm 7$% when FNDs were administered at 7 DIV (n=17 MEAs) and $47 \pm 8$ % when FNDs were administered at 14 DIV (n=16 MEAs).

On the contrary, the firing frequency was differently affected depending on the developmental stage of incubation with FNDs (7 versus 14 DIV). When FNDs were applied at 14 DIV (Figs. 3c and 3d), they drastically reduced the mean frequency on 18 DIV neurons with respect to controls (36%), from $1.03 \pm 0.07$ Hz to $0.66 \pm 0.05$ Hz ($p<0.001$). On the contrary, early incubation at 7 DIV preserved the firing frequency during culture maturation, as confirmed by measurements at 11 DIV ($0.23 \pm 0.03$ Hz for controls and $0.29 \pm 0.02$ Hz with FND, $p>0.05$) (Fig. 3e), at 14 DIV ($0.8 \pm 0.2$ Hz for controls and $0.60 \pm 0.04$ Hz with FND, $p>0.05$) (Fig. 3f) and at 18 DIV, with a mean firing frequency of $1.03 \pm 0.07$ Hz in control and $1.06 \pm 0.07$ Hz in the presence of FNDs (Fig. 3d).

Similarly, the mean number of bursts, evaluated over 120 s, decreased by 44%, (i.e. from $16 \pm 1$ to $9 \pm 1$, $p<0.001$), when FNDs were applied on mature networks, but remained unaltered when FNDs incubation occurred earlier (Fig. 3g). Burst duration was not significantly different among controls and FNDs (Fig. 3h).

Overall, these data suggest that spontaneous firing is significantly impaired only if the exposure to FNDs occurs at later stages of network maturation. Since early incubation (7 DIV) did not induce any significant effect, all other experiments were carried out by adding the FNDs at 14 DIV and recording the firing activity at 18 DIV.

Despite this reduction of spontaneous frequency, we could not find any significant difference on the degree of hippocampal network synchronization[30], independently of FNDs application at 7 or 14 DIV. This is clearly visible in Fig. 4, which shows the raster plots and the related cross-correlogram plotted for the different experimental conditions. In more details, the cross correlation probability in control neurons was $0.29 \pm 0.01$, and remained $0.28 \pm 0.01$ (7 DIV incubation) and $0.31 \pm 0.02$ (14 DIV incubations, p>0.1, Fig. 4b and 4c). Due to the role of GABAergic stimuli in triggering network synchronization[31-32], we next checked the effect of FNDs exposure on inhibitory and excitatory postsynaptic currents.

**The frequency of excitatory and inhibitory miniature postsynaptic currents is equally reduced by FNDs**

Analysis of miniature inhibitory and excitatory postsynaptic currents (mIPSCs and mEPSCs) revealed that exposure to FNDs at 14 DIV drastically reduced the mean frequency of spontaneous GABAergic (from $1.81 \pm 0.18$ Hz to $0.86 \pm 0.11$ Hz, p<0.01, n=5) and glutamatergic transmission (from $1.61 \pm 0.17$ Hz to $0.68 \pm 0.08$ Hz, p<0.01, n=5)[33]. Representative traces are shown in Figs. 5a and 5b, respectively for inhibitory and excitatory spontaneous events. Experiments were performed at 18 DIV. The mean frequency reduction was comparable for inhibitory and excitatory synapses (52% and 57% respectively, Fig. 5c). Unitary current amplitude peaked at $14.5 \pm 0.8$ pA and $12.4 \pm 0.3$ pA, respectively for inhibitory and excitatory spontaneous events, and these mean values were preserved after FNDs exposure, i.e. $14.2 \pm 0.6$ pA and $12.8 \pm 0.7$ pA, respectively (Fig. 5c). Similarly, other parameters such as the time-to-peak and the half-width of mPSCs remained unaltered upon FNDs application (p>0.1), as summarized in Fig. 5c. Overall, these findings suggest that, despite the marked frequency changes, GABAergic and glutamatergic transmission was equally affected during exposure to FNDs and thus likely preserving their balance.

**Resting membrane properties are preserved upon FNDs exposure**

Membrane input resistance ($R_{in}$) was measured to evaluate whether cell membrane integrity could be impaired by FNDs exposure. The protocol consisted in applying hyperpolarizing current pulses (from -100 pA to -120 pA in 10 mV steps), after holding the membrane resting potential at -70 mV and measuring the membrane voltage ($V_m$) deflection at the steady state (Fig. 6a). Representative traces for control neurons and those treated with FNDs are shown in Fig. 6a; $R_{in}$ was evaluated from the slope of the corresponding voltage-current relationship (Fig. 6b). Mean values of $R_{in}$ (inset) were in good agreement with previous values obtained from dissociated hippocampal neurons [34] and were not significantly altered by FNDs ($0.27 \pm 0.02$ G$\Omega$ (controls, p>0.1, n = 33) versus $0.25 \pm 0.02$ G$\Omega$ (FNDs, p>0.1, n = 26). Similarly, membrane resting potential ($V_{rest}$) at zero passing current (Fig. 6c) and rheobase (Fig. 6d) were not significantly perturbed by the nanoparticles. Mean values for $V_{rest}$ were - 54.2 ± 1.4 mV (control, p>0.1, n = 33) and -55.6 ± 3.1 mV (FNDs, p>0.1, n = 26). The rheobase, evaluated as the minimum amount of injected current necessary to induce AP firing when holding neurons at $V_h$ = -70 mV, was 124 ± 10 pA for controls cells and 150 ± 12 pA for FND-treated neurons (p>0.1)[35].

**Balance of fast-spiking versus slow-spiking neurons in cultured hippocampal neurons is mantained following exposure to FNDs**

Within the hippocampus, different populations of GABAergic interneurons tune the activity of pyramidal cells. These distinct phenotypes have been characterized in hippocampal slices on the basis of their electrophysiological properties and *post hoc* morphology[36,37]. Here we investigated this etherogeneity in dissociated hippocampal neurons. To this purpose, we measured in current-clamp configuration the firing properties of hippocampal neurons (18 DIV) by applying consecutive current pulses of increasing amplitude (from 100 pA to 300 pA, 10 pA steps, as described in Methods). As shown in Fig. 7a, the firing frequency increased with higher current injections; nevertheless, for some neurons the maximum value did not exceed 10 Hz (7.5 ± 1.4 pA for controls and 8.0 ± 1.3 pA for FNDs), whereas in other cases, it exceeded 20 Hz (23.5 ± 1.7 pA for controls and 25 ± 2 pA for FNDs). Thus, as reported for hippocampal slices[38], we could identify two different groups of neurons on the basis of their firing frequency, named as "slow-spiking" (SS) and "fast-spiking" (FS) neurons. Representative traces of (SS) and (FS) neurons are shown in Fig. 7b, either without FNDs (control, CTRL) or after FNDs treatment. AP firing patterns clearly differ among (SS) and (FS) neurons, rather than among controls versus neurons exposed to FNDs.

Slow and fast spiking neurons also differ for other aspects. First af all, they are characterized by different cell density, being the (SS) neurons more frequently recorded than the (FS) neurons. These latter indeed represent a minority of cells, namely 26% for controls and 38% for FNDs-treated cells. This may suggest that under our experimental conditions (dissociated neurons), FNDs do not significantly alter the balance among neurons subpopulations or do not specifically affect the activity of either neuron subpopulation.

(SS) and (FS) neurons also exhibit distinct $R_{in}$, in good agreement with the values reported for similar hippocampal neuronal phenotypes[36]. SS neurons had lower $R_{in}$ than (FS) neurons (respectively 0.25 GΩ versus 0.38 GΩ, p<0.05 for controls) and the same difference was mantained in the presence of FNDs (0.26 GΩ versus 0.32 GΩ, p<0.05, Fig. 7d). It is worth noting that the mean $R_{in}$ value, obtained by pooling all neurons together (as reported above in Fig. 6a), is effectively balanced among the two subpopulations.

Considering the action potential parameters, we found that in (SS) neurons the AP half width was higher than in (FS) neurons, i.e. 2.5 ms vs. 1.9 ms (p<0.05) and this difference was mantained after FNDs exposure: 2.3 ms vs. 1.9 ms (p<0.05, Fig. 7d). On the contrary, the threshold of activation (AP threshold, Fig. 7d), as well as the AP amplitude (not shown), were the same in all conditions (i.e. with or without FND, both for (SS) and (FS) neurons). A further difference among (SS) and (FS) neurons is represented by the firing adaptation, evaluated from the ratio of the istantaneous frequency measured at the beginning versus the one measured at the end of the AP train. The onset frequency ($f_o$) was estimated between the first and second AP, whereas the steady-state frequency ($f_{ss}$) corresponds to the firing frequency measured between the last two APs. Thus, from our data we can argue that fast spiking neurons were strongly adapting, being the decrease in firing frequency during the 2 s pulse equal to 40 ± 5 % for FNDs and 33 ± 7 % for controls (p<0.05), while adaptation was not significant in slow spiking neurons (p>0.1). Neuronal adaptation is correlated with the ability of neurons to synchronize with the surrounding network [39,40]. Again, these adaptation properties remained unaltered after FNDs seeding, in good agreement with the comparable cross correlation probability values measured by means of MEAs (Fig. 4).

Finally, another relevant difference among (SS) and (FS) neurons was observed using the phase-plot analysis (Fig. 8), which represents the time derivative of AP voltage (dV/dt) plotted versus voltage [41]. In this analysis we found that the maximum dV/dt value, associated with the density of voltage-gated $Na^+$ channels $Na_v$ [35,41], was significantly lower in (SS) with respect to (FS) neurons (115 versus 150 mV/ms, p<0.05), and, interestingly, the same difference was mantained after FNDs exposure (107 versus 145 mV/ms, p<0.05, Fig. 7d). From these data we can argue that the contribution of $Na_v$ currents to slow and fast-spiking neurons is preserved upon FNDs exposure.

**Action potential shape is altered by FNDs**
Concerning the action of FNDs on the single action potential shape, we found that FNDs altered the AP shape by drastically reducing the percentage of cells exhibiting the "kink", visualized as an abrupt voltage transition seen at the spike onset recorded from the soma [42]. The kink reflects the initiation of the spike in the initial segment of the neuron and is associated to the size and geometry of the axon initial segment and dendritic arborization. The kink, monitored in 40% of controls but only in 15% of FNDs-treated neurons, was proportionally distributed among SS and FS neurons (Fig. 8c). Using phase-plane plots, the kink can be visualized as the appearance of two distinct components reflecting the AP initiation in the axon initial segment and the subsequent somato-dendritic compartment[41]. Representative phase-plot graphs (with and without kink) are respectively shown in Figs. 8a and 8b, for both control and FNDs-treated neurons. The rapid AP onset was found in most of the controls, but was less evident in FNDs-treated neurons. Since the shape of the AP onset at the axon initial segment is accelerated in neurons with larger dendritic surface area[43], this suggests that a reduced geometry of the dendritic compartment, as in the case of FNDs exposure[25], or an impaired neurite extension, can be responsible for the slower spike onset and attenuation of the kink in FNDs-treated neurons. In addition, also an altered geometry, as the one induced by FNDs, can cause a reduced resistivity of the critical coupling between the axonal initial segment and soma[44].

It is worth noting that the FNDs-induced reduction of neurons showing with a fast onset of the AP could also be due to a different contribution of voltage gated $Na^+$ or $Ca^{2+}$ channels, caused by altered resting membrane potentials. However, we could not observe any significant difference among the resting potential of neurons with and without kink , in the presence or absence of FNDs ($p>0.1$ following ANOVA Bonferroni test).

**ODMR towards bio-sensing applications**

With the purpose of performing a preliminary feasibility test, we applied the Optically Detected Magnetic Resonance (ODMR)[45] technique to detect signals from FNDs incorporating NV centers that were internalized into hippocampal neurons. Negatively charged nitrogen-vacancy ($NV^-$) color centers in diamond have been recently proposed employed as an effective system for nanoscale magnetic field sensing in different contexts, including biological measurements in vitro[4,7,9,17], on the basis of their excellent magnetic field sensitivity and nano-scale size allowing for high spatial resolution. In particular, each $NV^-$ is can be regarded as an atomic-sized magnetic field sensor that can be optically read-out optically using ODMR.

Following the protocol described in Methods (Fig. 9a), we successfully performed a demonstrative ODMR measurement in a sample of hippocampal neurons hosting $NV^-$ centers in nanodiamonds. During an ODMR test, the sample is continuously illuminated and irradiated by microwaves at a frequency not far from the spin resonance. The frequency is slowly scanned while the photoluminescence (PL) is monitored. When approaches the resonance frequency, the photoluminescence will be diminished and a dip corresponding to the transition frequency (2.88 GHz) between the two spin states can be observed (Fig. 9b). Under these experimental conditions, spontaneous firing was preserved, as shown in Fig. 9c for two representative cells, respectively without (CTRL) and after FND treatment, suggesting that cell functionality was maintained after the implementation of the ODMR measurement protocol (ND hosting, mW-regime laser exposure, microwave excitation, as detailed in Methods).

**DISCUSSION**

Combining patch-clamp electrophysiology to detect AP and mPSC together with microelectrode arrays recordings, to monitor extracellular AP from developing neuronal networks, we investigated the functional effects of fluorescent diamond nanocrystals on primary cultures of hippocampal neurons. It is worth noting that, despite the outstanding optical properties and lack of cytotoxicity of FNDs have been largely explored[1-2], little is known regarding the effects of FNDs seeding on neuronal activity.

Here we show that FNDs induce a marked reduction of bursts firing frequency on reconstituted hippocampal networks and reduces the percentage of neurons displaying an abrupt AP onset phase (kink). Concerning the effect on hippocampal networks, we demonstrate that the stage of FNDs seeding (7 versus 14 DIV) is critical to modify the spontaneous firing frequency, causing a drastic reduction only if nanoparticles are applied at later stages of neuronal development. This functional impairment is in good agreement with previous findings reporting that, although FNDs do not exhibit cytotoxicity, they drastically alter the network morphology[25]. More specifically, the smaller the nanodiamond size, the longer the neurites extend [19].

Reduction of spontaneous firing frequency can be ascribed to a variety of pre- and post-synaptic factors: 1) the reduction of network excitability associated with a selective $Na^+$ or $Ca^{2+}$ channels inhibition or $K^+$ channels up-regulation[46], 2) the alteration of glutamatergic versus GABAergic transmission and the balance of fast-spiking versus slow-firing neurons[47], and 3) the modification of neuron morphology at the neuritic level[43,44]. To address these alternative possibilities, basic synaptic properties were measured by patch-clamp recordings. The frequency of both mIPSCs and mEPSCs events are equally reduced by FNDs, while their unitary amplitude and time course are preserved, suggesting an altered presynaptic activity induced by FND[48], although we cannot exclude a postsynaptic effect. This is in agreement with an impaired neurite growth observed in central and peripheral primary neurons without inducing cytotoxicity[25]. However, since the depressive effect on glutamatergic and GABAergic inputs are comparable, it is reasonable to hypothesize an "indirect" aspecific damage to the network rather than an effect of FNDs on specific cellular targets. As a direct consequence of reduced GABAergic and glutamatergic activity, the network burst firing frequency is halved. Despite the reduced excitability, it appears that FNDs do not compromise the network synchronization, in good agreement with the preservation of the inhibitory-excitatory balance and adaptation properties[49,50].

Regarding the AP generation in isolated neurons, it is relevant to underline how, even in the presence of FNDs and their aggregates, hippocampal neurons are still able to fire APs with unaltered input resistance, rheobase, firing frequency and AP waveform (Figs. 7 and 8), suggesting little or no effects to the gating and expression density of the ion channels sustaining the regular AP firing. Nevertheless, our data reveal an altered dynamics of the AP initiation induced by FNDs internalization. In particular, the abrupt onset of APs, which appears as a sharp kink detected in the majority of untreated neurons, is lacking in most of FND-treated neurons. Sharpening of the spike onset was originally attributed to a cooperation of $Na^+$ channels at the axonal initial segment[51], and later to the backpropagation of the AP from the axonal initial segment toward the soma[52]. A different interpretation, based on the "compartmentalization" hypothesis [42], attributes the kink to the distal initiation and the current sink caused by the different size of the soma and axon. In our work the first hypothesis cannot be confirmed by experimental findings, since the $Na^+$ channels density at the soma is preserved by FNDs (Fig. 8). Apart from that, the presence of the kink in the isolated hippocampal neurons used in our study reflects a good morphologic structure of the axonal initial segment[44] and the existence of a sufficiently well-extended dendritic surface area[43] that may be comparable to the AP recording in CA3 neurons of hippocampal slices with well-preserved morphology[53]. On the contrary, dendritic disruption (possibly associated with impaired neurite outgrowth) indeed alters the AP shape, even if it does not inhibit the firing itself. Thus, a reasonable hypothesis could be that FNDs and their aggregates alter the geometrical shape of hippocampal neurons *in vitro*[25] by limiting the extension of the neuronal branches and, consequently, reducing the network firing frequency. This is a critical issue worth to be considered in the perspective of applying FNDs for *in vivo* measurements.

Finally, we successfully implemented a test ODMR detection scheme to NDs internalized in hippocampal neurons demonstrating that these cells are not affected by the implementation of the measurement protocol, which (apart from the NDs internalization) is based on the application of both MW fields and laser illumination. Since each $NV^-$ centers can be regarded as atomic-sized magnetic field sensors that can be read-out optically using ODMR, in perspective their excellent magnetic field sensitivity can be applied to detect action potential propagation in 2D along the entire neuron with high-time resolution[4].

In conclusion, our findings give strong support to the use of FNDs for in vivo imaging and targetable drug delivery. The nanoparticles aggregation problem remains, in our opinion, the main cause of the observed alteration both at the single-cell level and in neuronal networks. Once solved this issue, FNDs could be more widely employed to "view" neuronal excitability using optical tools. It would be also useful to perform *in vivo* studies of FNDs delivery over longer periods of time than those used here, in order to exclude any possible long-term cytotoxicity due to the non-biodegradable nature of NDs.

## METHODS

### Cell cultures

All experiments were performed in accordance with the guidelines established by the National Council on Animal Care and approved by the local Animal Care Committee of Turin University. Hippocampal neurons were obtained from black-six mouse 18-day embryos. Hippocampus was rapidly dissected, kept in cold HBSS (4ºC) with high glucose, and digested with papain (0.5 mg/ml) dissolved in HBSS plus DNAse (0.1 mg/ml) [30,48,54]. Isolated cells were then plated at the final density of 1800 cells/mm$^2$ onto the MEA and 400 cells/mm$^2$ on 35 mm dishes (previously coated with poly-DL-lysine and laminine). Cells were incubated with 1% penicillin/streptomycin, 1% glutamax, 2.5% fetal bovine serum, 2% B-27 supplemented neurobasal medium in a humidified 5% $CO_2$ atmosphere at 37 °C. For MEA recordings, each MEA dish was covered with a fluorinated ethylene-propylene membrane (ALA scientific, Westbury, NY, USA) to reduce medium evaporation and maintain sterility, thus allowing repeated recordings from the same chip.

### FND preparation

The sample under exam is produced by ElementSix$^{TM}$ (UK) and consists of a synthetic diamond powder produced by disaggregation of High Pressure High Temperature (HPHT) single-crystals. The size of the diamond nanoparticles nominally ranges between ~10 nm and ~250 nm. The crystals are classified as type Ib having a nominal concentration of single substitutional nitrogen comprised between 10 ppm and 100 ppm.

Crystals also present impurities (superficial graphitic layers) and contaminants adsorbed on the surface, which are by-products of the fragmentation process. A preliminary cleaning process was therefore necessary to remove these impurities from crystals surface.

NDs were dispersed in an acid solution containing $H_2SO_4$ and $HNO_3$ (volumes ratio 9:1) for 72 hours at 75 °C. After filtration, nano-crystals still exhibited an acid behavior, and were therefore neutralized by means of a NaOH bath (2 hours, 90 °C). Processed powders were deposited over a suitable substrate for ion irradiation (1×1 cm$^2$ squares) creating an uniform layer with a thickness of ~20 μm (see fig 1b).

Since, due to the low amount of vacancies in the pristine powder, only a few percent of the substitutional nitrogen present in the diamond is involved in the formation of NV centers, ion-induced damaging is necessary to increase the vacancy density and therefore the total amount of luminescent NV centers upon subsequent thermal activation.

The samples were implanted with a 2 MeV H$^+$ ion broad beam at the AN2000 accelerator facility of the INFN Legnaro National Laboratories (INFN-LNL). An implantation fluence of $5\times10^{15}$ cm$^{-2}$ was delivered, guaranteeing the introduction of a vacancy density of $5\times10^{18}$ cm$^{-3}$ that is sufficiently far from the graphitization threshold of implanted diamond[55]. As resulting from the SRIM[56], Monte Carlo code simulation reported in Fig.1a, the penetration depth of 2 MeV proton is sufficient to cross through the thin (~20 µm) film of nanodiamond powder, guaranteeing a nearly-uniform vacancy creation over the totality of the nano-particles. Thermal treatment (800 °C for 1 h in nitrogen environment) was performed after ion implantation to promote the formation of NV$^-$ centers. Assuming a NV formation efficiency of ~10%[26], the described processing protocol guarantees that even the smaller nano-particles, i.e. the ones that can be easily uptaken by cells, contain several tens of NV color centers.

**Microelectrode arrays recordings**

Multisite extracellular recordings were carried out with the MultiChannel System MCS, (Reutlingen Germany). Data acquisition was controlled through the MC_Rack software, by setting the threshold for spike detection at -30 µV and sampling at 10 kHz. Recording duration was 2 minutes. Burst analysis was performed using Neuroexplorer software (Nex Technologies, Littleton MA USA) after spike sorting. A burst is defined as a group of spikes with decreasing amplitude [41], thus we set a threshold of at least 3 spikes and a minimum of 10 ms duration. We set interval algorithm specifications such as maximum interval to start burst (0.17 s) and maximum interval to end burst (0.3 s) recorded in 0.02 s bins. Burst analysis was conducted considering two main parameters: mean frequency and number of bursts. In order to examine synchronicity, cross-correlation probability vs. time diagrams were constructed by means of Neuroexplorer software, using ± 0.5 s and 5 ms bin size. Raster plots were created by using the specific function on Neuroexplorer software.

**Cell incubation with FNDs and confocal imaging**

Hippocampal neurons incubation with FNDs was performed at 7 DIV or 14 DIV, as specified in the Results. For both MEAs and plastic dishes, half of the culture medium (i.e. approximately 1 ml) was replaced with FND-medium to reach a final concentration of 40 µg/ml. Incubation with the cytoplasmic membrane dye was performed 24 hours later. For confocal imaging, hippocampal neurons were plated on 35 mm dishes (ibidi GmbH, Planegg/ Martinsried, Germany).

Imaging was performed using a Leica TCS SP5-AOBS 5-channel confocal system (Leica Microsystems) equipped with an argon ion and a 561nm DPSS laser. Cells were imaged using a HCX PL APO 63x/1.4 NA oil immersion objective at a pixel resolution of 0.08x0.08x0.3μm.

The luminescent emission from the FNDs was excited by 561 nm laser, while the emission was collected in the 650 - 750 nm spectral range. The same excitation wavelength was uneffective in untreated neurons. Green fluorescence for intracellular staining was obtained by 488 nm wavelength. Image analysis was performed using ImageJ software.

**Patch-clamp experiments**

Patch-clamp experiments were performed using pCLAMP software (Axon Instruments, Foster City, CA) and a 12-bit A/D Tecmar Laboratory Master board (125 kHz). All the experiments were performed at room temperature (22 – 24°C). Data analysis was performed with CLAMPFIT (Axon Instruments, Foster City, CA).

Voltage-clamp experiments

Miniature postsynaptic currents (mPSCs) were recorded under voltage-clamp conditions from mature hippocampal neurons (18 DIV) as detailed in [46,48,54,57]. For both controls and FND-incubated cells, holding membrane potential ($V_h$) was maintained at -70 mV and data acquired for two minutes.

Miniature postsynaptic currents were acquired with sample frequency ranging between 10 kHz and filtered at half the acquisition rate with an 8-pole low-pass Bessel filter. Recordings with leak currents ≥ 100 pA or series resistance ≥ 20 MΩ were discarded. Miniature inhibitory and excitatory currents were recorded by superfusing the whole-cell clamped postsynaptic neuron with Tyrode solution containing (in mM): 2 $CaCl_2$, 150 NaCl, 1 $MgCl_2$, 10 Hepes, 10 glucose, 4 KCl (pH 7.4). Tetrodotoxin (0.3 μM) (Tocris Cookson Ltd, Bristol, UK) was added to block voltage dependent $Na^+$ channels and spontaneous action potentials propagation. GABAergic and glutamatergic currents were respectively isolated by D-AP5 (50 μM) and Picrotoxin (100 μM). The standard internal solution contained (in mM): 90 CsCl, 20 TEACl, 10 EGTA, 10 glucose, 1 $MgCl_2$, 4 ATP, 0.5 GTP and 15 phosphocreatine (pH 7.4). For unitary mini analysis, all double or multiple events were discarded[48,54]. The rheobase was evaluated as the minimum amount of injected current necessary to induce AP firing, when holding neurons at $V_h$ = -70 mV (see [35] for details)

Current-clamp experiments

Intracellular solution for current-clamp experiments contained (in mM): 135 gluconic acid (potassium salt: K-gluconate), 5 NaCl, 2 $MgCl_2$, 10 HEPES, 0.5 EGTA, 2 ATP-Tris, 0.4 Tris-GTP[46]. These trials were performed in physiological Tyrode saline solution enriched with kynurenic acid (1 mM) and picrotoxin (100 µM). Action potential firing was evoked by applying 2 second current pulses injection (from 100 pA to 300 pA, 10 pA steps). Analysis of AP shape, phase-plane plot and data interpretation on AP trains were performed as previously detailed[35].

**ODMR**

The ODMR technique is based on the dependence of the photoluminescence (PL) emission intensity from $NV^-$ defects in diamond on the spin state of the defect, thus enabling different types of electro-magnetic detection schemes[5,7,17,45]. In our setup (Fig. 9a) a non-resonant 532 nm continuous laser excitation was employed to initialize the $NV^-$ centers in the bright $m_s$=0 spin state. Then, a microwave field of -20 dBm power was generated at variable frequency in the 2.80-2.96 GHz range, by positioning a micro-strip antenna in close proximity of the investigated sample. This enabled the observation of a local minimum in the PL emission intensity corresponding to the resonance frequency (2.88 GHz) of the ground-state zero-field splitting between the $|0\rangle$ and $|\pm 1\rangle$ states[45]. In this way we have exposed the cells to a measurement protocol that in perspective would enable detecting in a non-destructive way the weak magnetic field associated to the action potential firing[4].

**Statistical analysis**

Data are given as the mean ± SEM for the number (n) of cells. Statistical significance was estimated with paired Student's t tests in case two groups of measurements had to be compared and with a one-way ANOVA followed by post hoc Bonferroni analysis in case more than two groups had to be compared with one another.
Data were found statistically significant when p< 0.05 (*). Data analysis was performed with pClamp and Origin software (OriginLab Corporation, Northampton, MA, USA).

**Data availability statement**

The datasets generated during and/or analysed during the current study are available from the corresponding author on reasonable request.

# References

1. Xi, G. *et al.* Convection-enhanced delivery of nanodiamond drug delivery platforms for intracranial tumor treatment. *Nanomedicine : nanotechnology, biology, and medicine* **10**, 381-391, doi:10.1016/j.nano.2013.07.013 (2014).
2. Kaur, R. & Badea, I. Nanodiamonds as novel nanomaterials for biomedical applications: drug delivery and imaging systems. *International journal of nanomedicine* **8**, 203-220, doi:10.2147/ijn.s37348 (2013).
3. Zaitsev, A. M. Optical Properties of Diamond—Data Handbook doi:10.1007/978-3-662-04548-0 (2001).
4. Barry, J. F. *et al.* Optical magnetic detection of single-neuron action potentials using quantum defects in diamond. *Proceedings of the National Academy of Sciences* **113**, 14133-14138, doi:10.1073/pnas.1601513113 (2016).
5. Glenn, D. R. *et al.* Single-cell magnetic imaging using a quantum diamond microscope. *Nat Meth* **12**, 736-738, doi:10.1038/nmeth.3449 http://www.nature.com/nmeth/journal/v12/n8/abs/nmeth.3449.html#supplementary-information (2015).
6. Hall, L. T. *et al.* Monitoring ion-channel function in real time through quantum decoherence. *Proc Natl Acad Sci U S A* **107**, 18777-18782, doi:10.1073/pnas.1002562107 (2010).
7. Kucsko, G. *et al.* Nanometre-scale thermometry in a living cell. *Nature* **500**, 54-58, doi:10.1038/nature12373 http://www.nature.com/nature/journal/v500/n7460/abs/nature12373.html#supplementary-information (2013).
8. Posfai, M. & Dunin-Borkowski, R. E. Imaging: Magnetic bacteria on a diamond plate. *Nature* **496**, 442-443, doi:10.1038/496442a (2013).
9. McGuinness, L. P. *et al.* Quantum measurement and orientation tracking of fluorescent nanodiamonds inside living cells. *Nat Nano* **6**, 358-363, doi:http://www.nature.com/nnano/journal/v6/n6/abs/nnano.2011.64.html#supplementary-information (2011).
10. Hui, Y. Y. *et al.* Wide-field imaging and flow cytometric analysis of cancer cells in blood by fluorescent nanodiamond labeling and time gating. *Scientific reports* **4**, 5574, doi:10.1038/srep05574 (2014).
11. Hsu, T. C., Liu, K. K., Chang, H. C., Hwang, E. & Chao, J. I. Labeling of neuronal differentiation and neuron cells with biocompatible fluorescent nanodiamonds. *Scientific reports* **4**, 5004, doi:10.1038/srep05004 (2014).
12. Rehor, I. *et al.* Fluorescent Nanodiamonds Embedded in Biocompatible Translucent Shells. *Small* **10**, 1106-1115, doi:10.1002/smll.201302336 (2014).
13. Haziza, S. *et al.* Fluorescent nanodiamond tracking reveals intraneuronal transport abnormalities induced by brain-disease-related genetic risk factors. *Nature nanotechnology* **12**, 322-328, doi:10.1038/nnano.2016.260 (2017).
14. Igarashi, R., doi:DOI: 10.1021/nl302979d (2012).
15. Balasubramanian, G., Lazariev, A., Arumugam, S. R. & Duan, D. W. Nitrogen-Vacancy color center in diamond-emerging nanoscale applications in bioimaging and biosensing. *Curr. Opin. Chem. Biol.* **20**, 69-77, doi:10.1016/j.cbpa.2014.04.014 (2014).
16. Hsiao, W. W., Hui, Y. Y., Tsai, P. C. & Chang, H. C. Fluorescent Nanodiamond: A Versatile Tool for Long-Term Cell Tracking, Super-Resolution Imaging, and Nanoscale Temperature Sensing. *Accounts of chemical research* **49**, 400-407, doi:10.1021/acs.accounts.5b00484 (2016).
17. Hemelaar, S. R. *et al.* The interaction of fluorescent nanodiamond probes with cellular media. *Mikrochimica acta* **184**, 1001-1009, doi:10.1007/s00604-017-2086-6 (2017).

18 Hemelaar, S. R. *et al.* Nanodiamonds as multi-purpose labels for microscopy. *Scientific reports* **7**, 720, doi:10.1038/s41598-017-00797-2 (2017).
19 Edgington, R. J. *et al.* Patterned neuronal networks using nanodiamonds and the effect of varying nanodiamond properties on neuronal adhesion and outgrowth. *Journal of neural engineering* **10**, 056022, doi:10.1088/1741-2560/10/5/056022 (2013).
20 Thalhammer, A., Edgington, R. J., Cingolani, L. A., Schoepfer, R. & Jackman, R. B. The use of nanodiamond monolayer coatings to promote the formation of functional neuronal networks. *Biomaterials* **31**, 2097-2104, doi:10.1016/j.biomaterials.2009.11.109 (2010).
21 Mohan, N., Chen, C. S., Hsieh, H. H., Wu, Y. C. & Chang, H. C. In vivo imaging and toxicity assessments of fluorescent nanodiamonds in Caenorhabditis elegans. *Nano letters* **10**, 3692-3699, doi:10.1021/nl1021909 (2010).
22 Vaijayanthimala, V. *et al.* The long-term stability and biocompatibility of fluorescent nanodiamond as an in vivo contrast agent. *Biomaterials* **33**, 7794-7802, doi:10.1016/j.biomaterials.2012.06.084 (2012).
23 Liu, K. K., Wang, C. C., Cheng, C. L. & Chao, J. I. Endocytic carboxylated nanodiamond for the labeling and tracking of cell division and differentiation in cancer and stem cells. *Biomaterials* **30**, 4249-4259, doi:10.1016/j.biomaterials.2009.04.056 (2009).
24 Silbajoris, R. *et al.* Detonation nanodiamond toxicity in human airway epithelial cells is modulated by air oxidation. *Diamond and Related Materials* **58**, 16-23, doi:http://doi.org/10.1016/j.diamond.2015.05.007 (2015).
25 Huang, Y. A. *et al.* The effect of fluorescent nanodiamonds on neuronal survival and morphogenesis. *Scientific reports* **4**, 6919, doi:10.1038/srep06919 (2014).
26 Rabeau, J. R. *et al.* Single nitrogen vacancy centers in chemical vapor deposited diamond nanocrystals. *Nano letters* **7**, 3433-3437, doi:10.1021/nl0719271 (2007).
27 Vlasov, II *et al.* Nitrogen and luminescent nitrogen-vacancy defects in detonation nanodiamond. *Small* **6**, 687-694, doi:10.1002/smll.200901587 (2010).
28 Faklaris, O. *et al.* Detection of single photoluminescent diamond nanoparticles in cells and study of the internalization pathway. *Small* **4**, 2236-2239, doi:10.1002/smll.200800655 (2008).
29 Neugart, F. *et al.* Dynamics of diamond nanoparticles in solution and cells. *Nano letters* **7**, 3588-3591, doi:10.1021/nl0716303 (2007).
30 Gavello, D. *et al.* Leptin Counteracts the Hypoxia-Induced Inhibition of Spontaneously Firing Hippocampal Neurons: A Microelectrode Array Study. *Plos One* **7**, doi:10.1371/journal.pone.0041530 (2012).
31 Bonifazi, P. *et al.* GABAergic hub neurons orchestrate synchrony in developing hippocampal networks. *Science (New York, N.Y.)* **326**, 1419-1424, doi:10.1126/science.1175509 (2009).
32 Picardo, M. A. *et al.* Pioneer GABA cells comprise a subpopulation of hub neurons in the developing hippocampus. *Neuron* **71**, 695-709, doi:10.1016/j.neuron.2011.06.018 (2011).
33 Baldelli, P., Fassio, A., Valtorta, F. & Benfenati, F. Lack of synapsin I reduces the readily releasable pool of synaptic vesicles at central inhibitory synapses. *J Neurosci* **27**, 13520-13531, doi:10.1523/jneurosci.3151-07.2007 (2007).
34 Fan, J. *et al.* Reduced Hyperpolarization-Activated Current Contributes to Enhanced Intrinsic Excitability in Cultured Hippocampal Neurons from PrP(-/-) Mice. *Frontiers in cellular neuroscience* **10**, 74, doi:10.3389/fncel.2016.00074 (2016).
35 Vandael, D. H., Zuccotti, A., Striessnig, J. & Carbone, E. Ca(V)1.3-driven SK channel activation regulates pacemaking and spike frequency adaptation in mouse chromaffin cells. *J Neurosci* **32**, 16345-16359, doi:10.1523/jneurosci.3715-12.2012 (2012).
36 Bjorefeldt, A., Wasling, P., Zetterberg, H. & Hanse, E. Neuromodulation of fast-spiking and non-fast-spiking hippocampal CA1 interneurons by human cerebrospinal fluid. *The Journal of physiology* **594**, 937-952, doi:10.1113/jp271553 (2016).
37 Lorincz, T., Kisfali, M., Lendvai, B. & Sylvester Vizi, E. Phenotype-dependent Ca(2+) dynamics in single boutons of various anatomically identified GABAergic interneurons in the rat hippocampus. *The European journal of neuroscience* **43**, 536-547, doi:10.1111/ejn.13131 (2016).

38 Schaub, C., Uebachs, M., Beck, H. & Linnebank, M. Chronic homocysteine exposure causes changes in the intrinsic electrophysiological properties of cultured hippocampal neurons. *Experimental brain research* **225**, 527-534, doi:10.1007/s00221-012-3392-1 (2013).

39 Fuhrmann, G., Markram, H. & Tsodyks, M. Spike frequency adaptation and neocortical rhythms. *Journal of neurophysiology* **88**, 761-770 (2002).

40 Ladenbauer, J., Augustin, M., Shiau, L. & Obermayer, K. Impact of adaptation currents on synchronization of coupled exponential integrate-and-fire neurons. *PLoS computational biology* **8**, e1002478, doi:10.1371/journal.pcbi.1002478 (2012).

41 Bean, B. P. The action potential in mammalian central neurons. *Nature reviews. Neuroscience* **8**, 451-465, doi:10.1038/nrn2148 (2007).

42 Brette, R. Sharpness of spike initiation in neurons explained by compartmentalization. *PLoS computational biology* **9**, e1003338, doi:10.1371/journal.pcbi.1003338 (2013).

43 Eyal, G., Mansvelder, H. D., de Kock, C. P. & Segev, I. Dendrites impact the encoding capabilities of the axon. *J Neurosci* **34**, 8063-8071, doi:10.1523/jneurosci.5431-13.2014 (2014).

44 Telenczuk, M., Fontaine, B. & Brette, R. The basis of sharp spike onset in standard biophysical models. *PLoS One* **12**, e0175362, doi:10.1371/journal.pone.0175362 (2017).

45 Gruber, A. *et al.* Scanning Confocal Optical Microscopy and Magnetic Resonance on Single Defect Centers. *Science (New York, N.Y.)* **276**, 2012-2014, doi:10.1126/science.276.5321.2012 (1997).

46 Gavello, D. *et al.* Early Alterations of Hippocampal Neuronal Firing Induced by Abeta42. *Cerebral cortex (New York, N.Y. : 1991)*, doi:10.1093/cercor/bhw377 (2016).

47 Palop, J. J. & Mucke, L. Amyloid-beta-induced neuronal dysfunction in Alzheimer's disease: from synapses toward neural networks. *Nature neuroscience* **13**, 812-818, doi:10.1038/nn.2583 (2010).

48 Baldelli, P., Hernandez-Guijo, J. M., Carabelli, V. & Carbone, E. Brain-derived neurotrophic factor enhances GABA release probability and nonuniform distribution of N- and P/Q-type channels on release sites of hippocampal inhibitory synapses. *Journal of Neuroscience* **25**, 3358-3368, doi:10.1523/jneurosci.4227-04.2005 (2005).

49 Marcantoni, A., Raymond, E. F., Carbone, E. & Marie, H. Firing properties of entorhinal cortex neurons and early alterations in an Alzheimer's disease transgenic model. *Pflugers Archiv : European journal of physiology* **466**, 1437-1450, doi:10.1007/s00424-013-1368-z (2014).

50 Benda, J. & Herz, A. V. A universal model for spike-frequency adaptation. *Neural computation* **15**, 2523-2564, doi:10.1162/089976603322385063 (2003).

51 Naundorf, B., Wolf, F. & Volgushev, M. Unique features of action potential initiation in cortical neurons. *Nature* **440**, 1060-1063, doi:10.1038/nature04610 (2006).

52 Yu, Y., Shu, Y. & McCormick, D. A. Cortical action potential backpropagation explains spike threshold variability and rapid-onset kinetics. *J Neurosci* **28**, 7260-7272, doi:10.1523/jneurosci.1613-08.2008 (2008).

53 Ho, E. C., Struber, M., Bartos, M., Zhang, L. & Skinner, F. K. Inhibitory networks of fast-spiking interneurons generate slow population activities due to excitatory fluctuations and network multistability. *J Neurosci* **32**, 9931-9946, doi:10.1523/jneurosci.5446-11.2012 (2012).

54 Baldelli, P., Novara, M., Carabelli, V., Hernandez-Guijo, J. M. & Carbone, E. BDNF up-regulates evoked GABAergic transmission in developing hippocampus by potentiating presynaptic N- and P/Q-type Ca2+ channels signalling. *European Journal of Neuroscience* **16**, 2297-2310, doi:10.1046/j.1460-9568.2002.02313.x (2002).

55 Battiato, A. *et al.* Softening the ultra-stiff: Controlled variation of Young's modulus in single-crystal diamond by ion implantation. *Acta Materialia* **116**, 95-103, doi:http://dx.doi.org/10.1016/j.actamat.2016.06.019 (2016).

56 Ziegler, J. F., Ziegler, M. D. & Biersack, J. P. SRIM - The stopping and range of ions in matter (2010). *Nucl. Instrum. Methods Phys. Res. Sect. B-Beam Interact. Mater. Atoms* **268**, 1818-1823, doi:10.1016/j.nimb.2010.02.091 (2010).
57 Allio, A. *et al.* Bud extracts from Tilia tomentosa Moench inhibit hippocampal neuronal firing through GABAA and benzodiazepine receptors activation. *Journal of ethnopharmacology* **172**, 288-296, doi:10.1016/j.jep.2015.06.016 (2015).

**FIGURE LEGENDS**

**Figure 1. SEM imaging and SRIM simulation.** a) SEM micrograph of dispersed nanodiamonds over a silicon substrate: the mean crystal dimension is smaller than 100 nm. b) SEM micrograph of a 20 µm thick nanodiamond deposition before ion irradiation. c) SRIM Monte Carlo Simulation of the damage profile induced by 2 MeV protons over the diamond deposition: the ions cross the sample creating a quasi-uniform defect concentration over the whole depth. d) NV centre luminescence spectrum**.** Confocal fluorescence spectrum of FNDs. The NV- ZPL peak is clearly visible at 638 nm, as well as the corresponding wide phonon band at larger wavelengths.

**Figure 2. FND internalization into cultured hippocampal neurons.** a) Confocal fluorescence micrograph of cultured hippocampal neurons (14 DIV), exposed to 40 µg/ml FND for 2 days, and stained in green with the cytoplasmic labelling dye (CellTracker™ Green CMFDA, ThermoFisher). Red emission is from FNDs. The entire field and cross-sections (XZ and YZ) were shown.b) Hippocampal neurons were exposed to 40 µg/ml FND for for 4 and 11 days and stained as in a) at 18DIV. Maximum intensity projections of confocal z-stacks are shown. c) Quantification of FNDs internalized in neurons treated as in b). Scale bar represents 10 µm.

**Figure 3. MEA recordings of hippocampal neurons activity without FNDs and after FNDs seeding**. a-c) Representative traces of spontaneous firing at 18 DIV (data from 3 representative MEA channels) under control conditions (CTRL), without FNDs (a), with FNDs seeded at 7 DIV (b) and at 14 DIV (c). Insets: higher magnification of single spikes and bursts. d) Bar graphs of the mean frequency measured at 18 DIV without FNDs (CTRL, white) and after FNDs administration at 7 DIV (grey) and 14 DIV (black). e) Bar graphs of mean frequency of the spontaneous activity measured at 11 DIV without FNDs (CTRL) and after FNDs administration at 7 DIV. f) Bar graphs of mean frequency of the spontaneous activity measured at 14 DIV without FNDs (CTRL) and after FNDs administration at 7 DIV. g) Bar graphs of mean number of bursts and mean burst duration without FNDs (CTRL), and after FNDs administration at 7 and 14 DIV. h) Bar graphs of mean burst duration (s) without (CTRL) and after FND seeding at 7 and 14 DIV.

**Figure 4. Raster plots and cross-correlograms.** a) Representative raster plots showing the occurrence of the events in a representative control MEA (left) and those MEAs exposed to FNDs at 7 and 14 DIV, respectively (centre, right). b) Cross-correlogram plots showing the probability of coincidence of the events in control condition and after exposure to FNDs. c) Mean value of the cross-correlation peak in control condition (CTRL) and after exposure to FNDs at 7 DIV (grey) and 14 DIV (black).

**Figure 5. Frequency of miniature inhibitory and excitatory currents is reduced by FNDs.** a) Occurrence of miniature inhibitory and b) excitatory post synaptic currents is reduced by FNDs with respect to controls. Inset: averaged miniature currents in control (black) and after FND seeding (grey). c) Bar graphs showing the mean values of the following mPSCs parameters: frequency, unitary amplitude, half width and time to peak, for control (white bars) and FNDs-treated neurons (black bars).

**Figure 6. Passive membrane properties are not altered by FNDs.** a) Top: current-clamp protocol for measuring the membrane input resistance ($R_{in}$), consisting of consecutive current pulses (-100, -120 pA range) with 10 pA steps, 2 s duration. Bottom: representative traces under control conditions (CTRL) and with FNDs. b) $V_m$ deflections (measured from (a)) plotted versus current pulses. Inset: mean values of $R_{in}$, under control conditions (CTRL, white) and with FNDs (black). c) Mean values of membrane resting potential ($V_{rest}$) and d) mean values of rheobase in the two experimental conditions (control versus FND).

**Figure 7. Primary cultures of hippocampal neurons exhibit (SS) and (FS) neurons, independently of FNDs.** a) Firing frequency plotted versus increasing current amplitude pulses, applied in the 100-300 pA range at 10 pA steps. For both controls and neurons incubated with FNDs, frequencies increased linearly with applied current; though, most of the cells exhibited maximum firing frequency below 10 Hz, (SS), while a minority reached higher values (FS). b) Different adaptation among (SS) and (FS) neurons. Istantaneous frequency was evaluated at the onset ($f_o$) and at the steady-state ($f_{s-s}$) of the pulse. c) Bar graphs representing mean $f_o$ and $f_{s-s}$. Significant adaptation in frequency is monitored exclusively for (FS) neurons. d) (FS) and (SS) neurons exhibit different $R_{in}$, AP half width, maximum dV/dt. All these parameters are unaltered by FNDs.

**Figure 8. Altered AP onset by FND.** Representative AP and corresponding phase-plots for control cells (upper panels) and FND-treated (bottom). FND seeding reduced the number of cells exhibiting the kink from 40% to 15%.

**Figure 9. ODMR measurement.** a) Schematics of the confocal setup for ODMR measurements. The sample was observed via a single-photon-sensitive confocal microscope integrated to the Olimpus IX73, the excitation light being provided by a solid state laser at 532 nm. A dichroic beamsplitter (long-pass at 570 nm) reflected the excitation light (3 mW maximum) inside the air objective (Olympus, 60×, NA = 0.9) focusing inside the sample and transmits the photoluminescence towards the detecting apparatus. A microwave field with power $P_{mw}$ = -20 dBm was transmitted via a micro-strip antenna. The experimental control and data acquisition were performed using the open-source Qudi software suite. b) ODMR signal detected from NV centers in nanodiamonds internalized in a neural cell. c) Spontaneous firing of hippocampal neurons, without and with FNDs, is preserved by the application of the ODMR protocol.

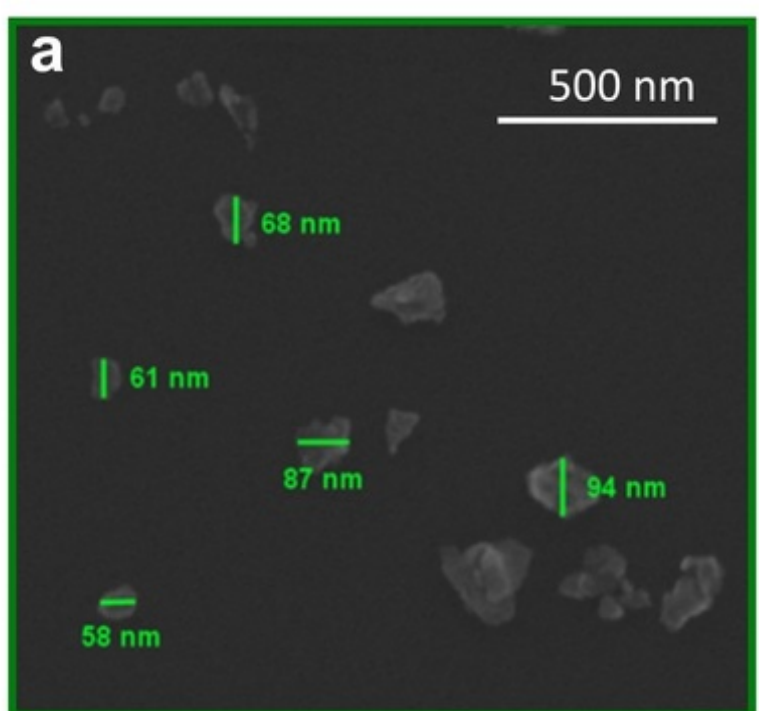
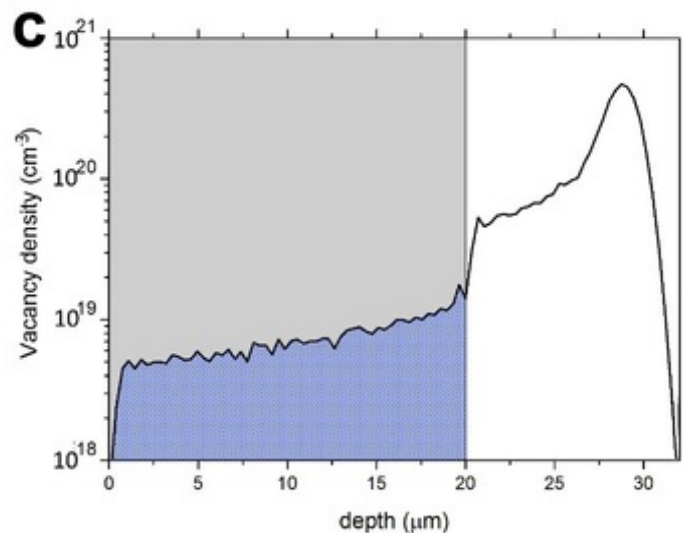
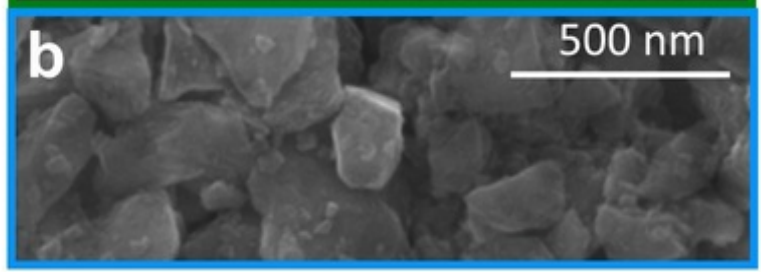
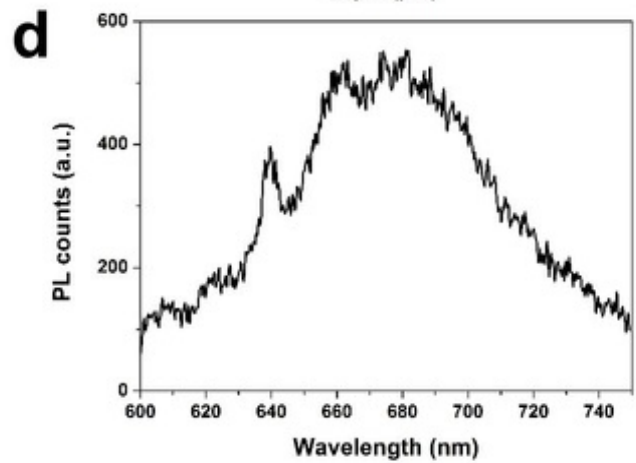

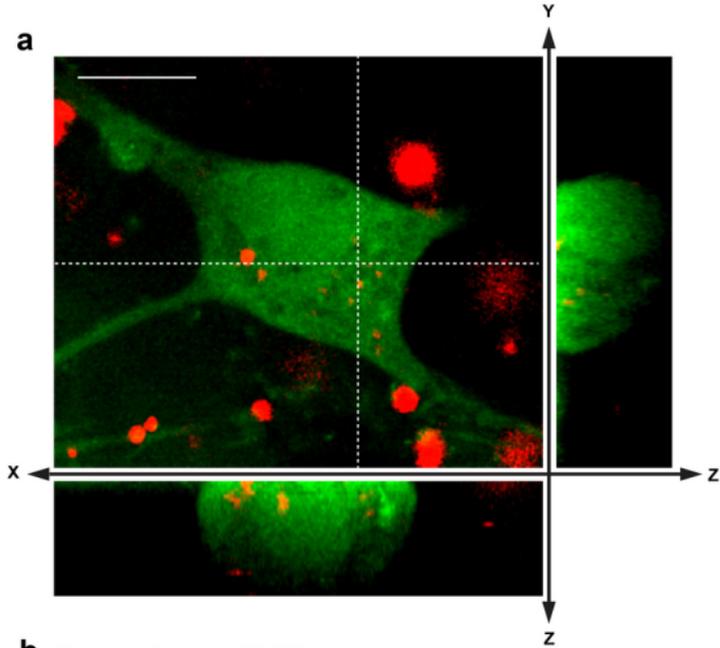

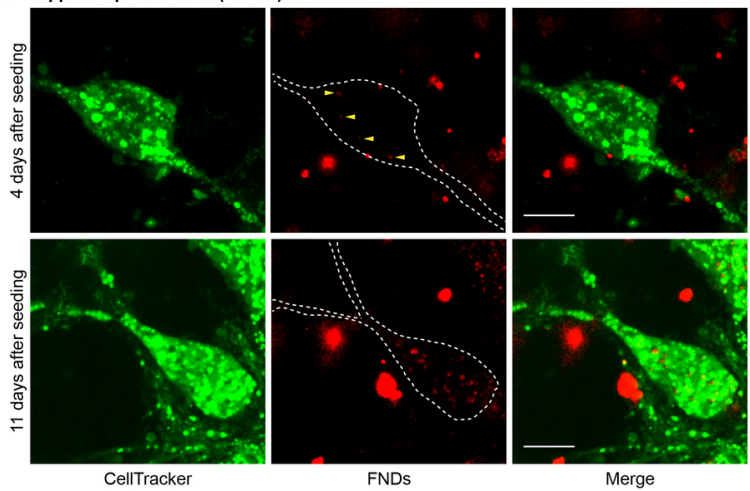

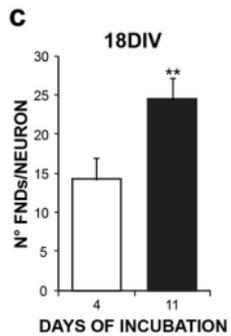

CellTracker    FNDs    Merge

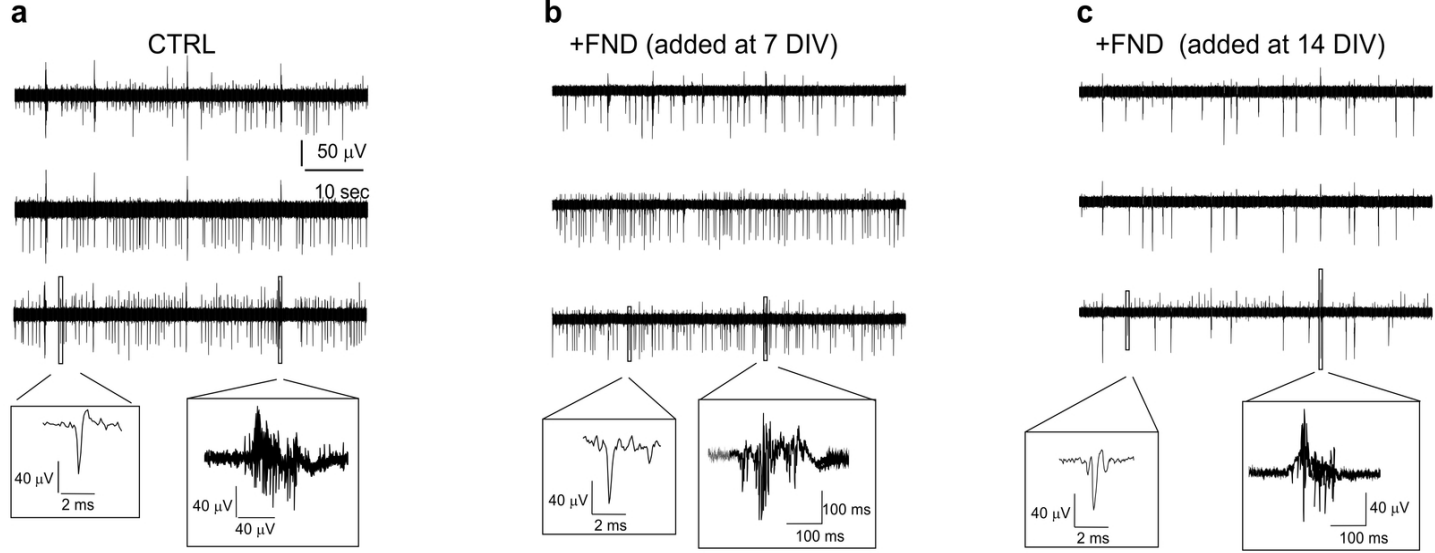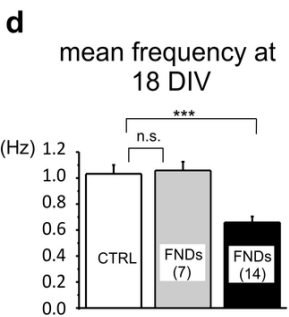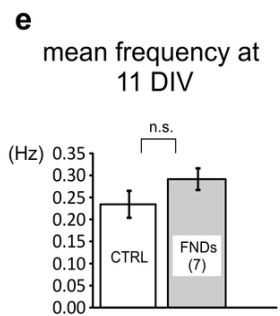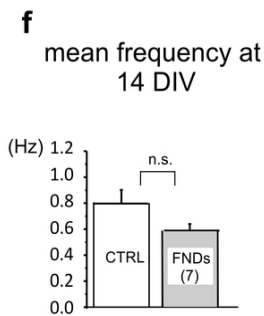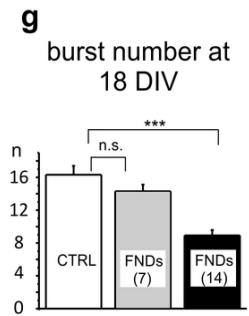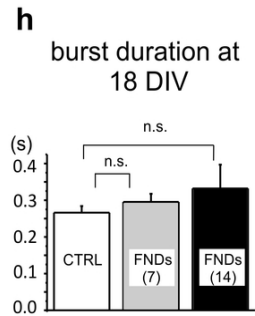

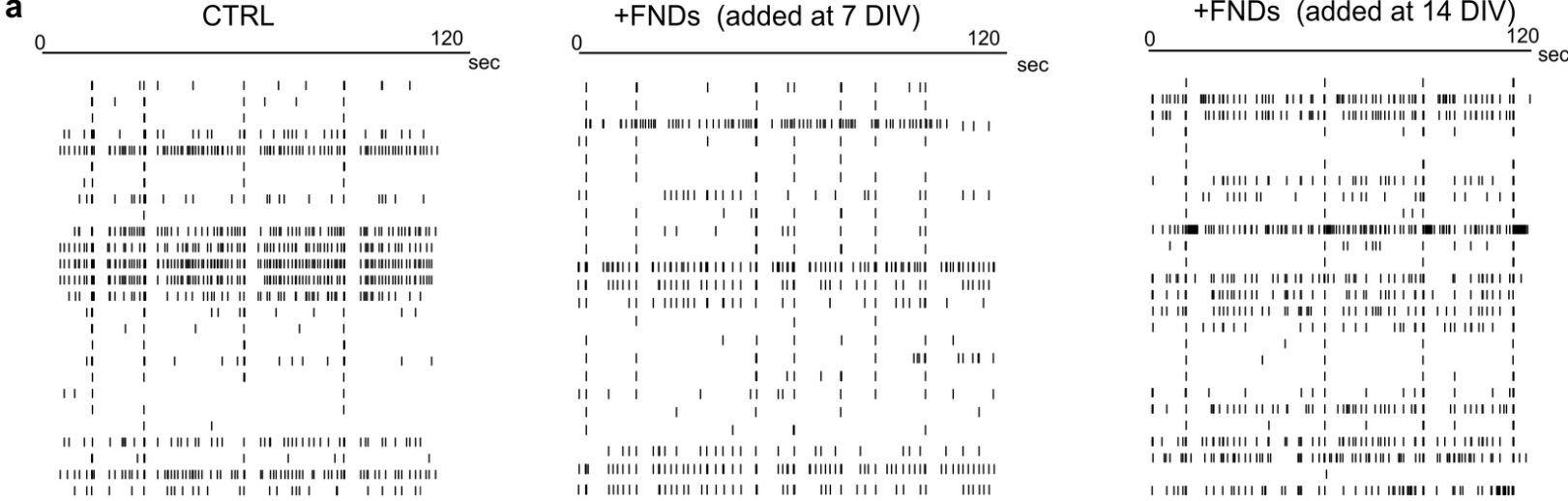

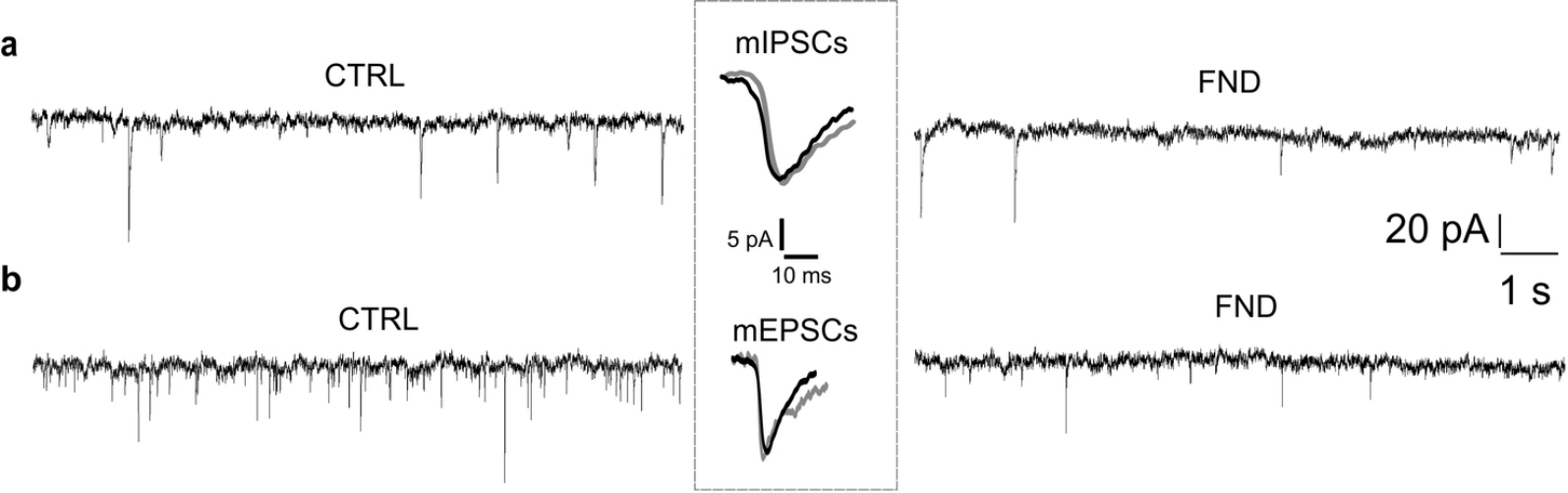
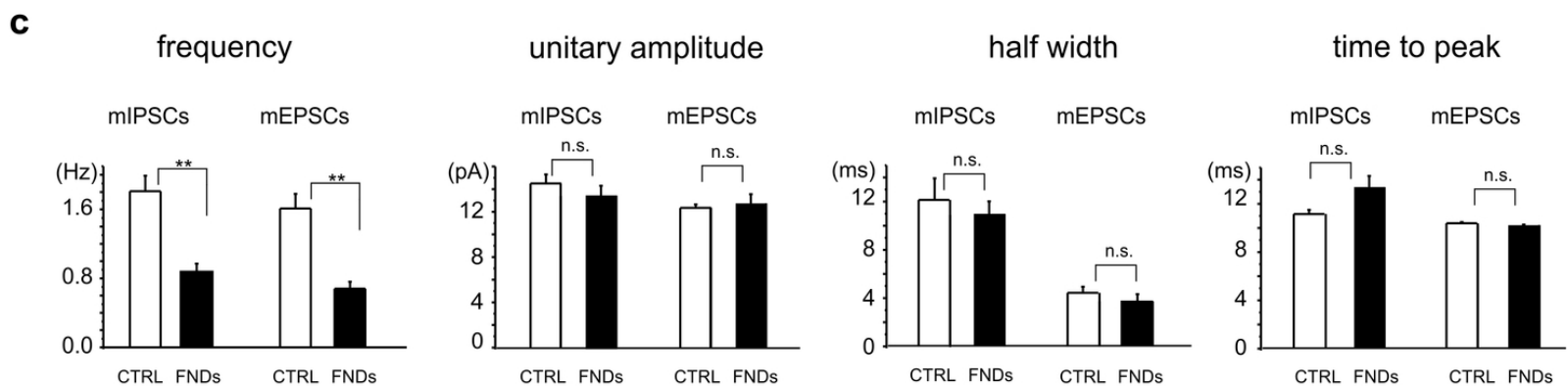

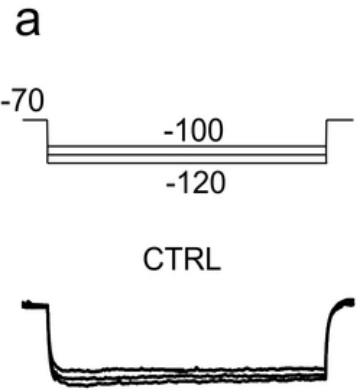
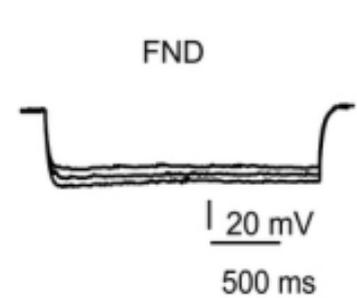
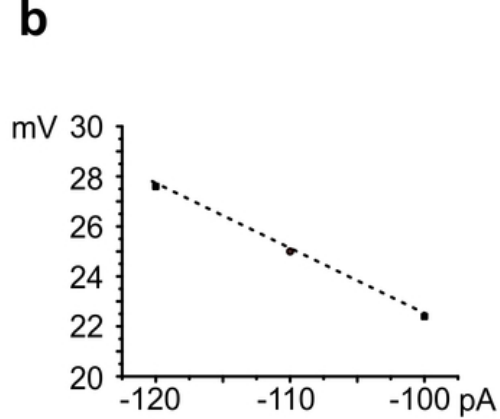
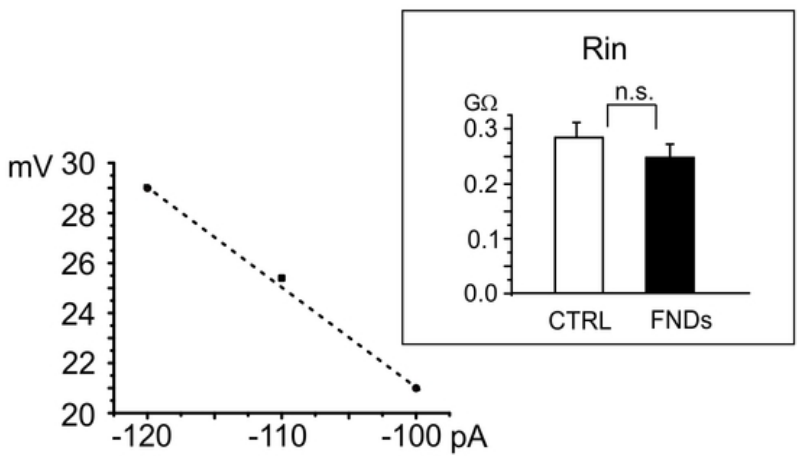
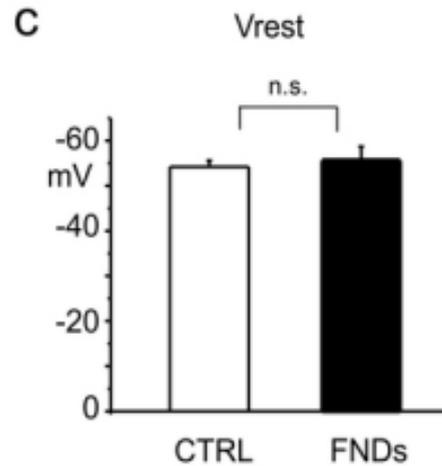
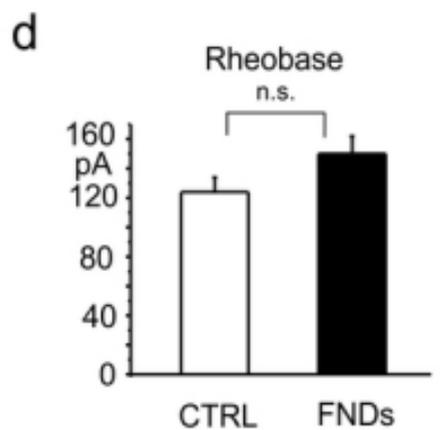

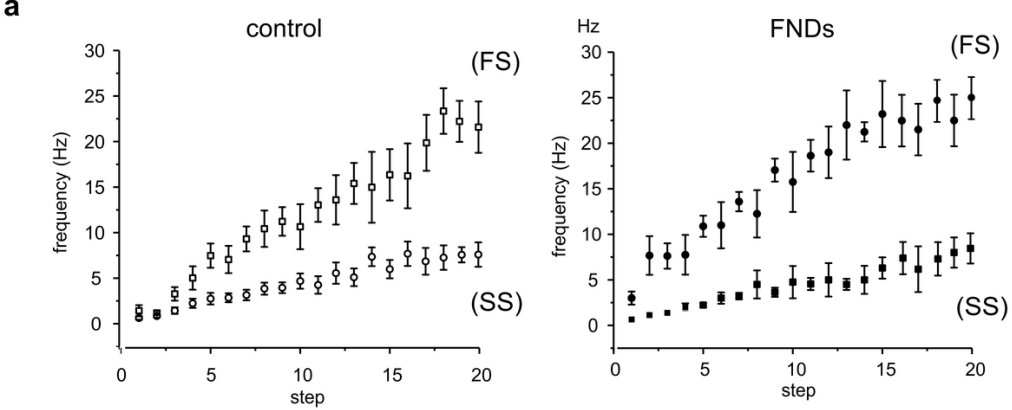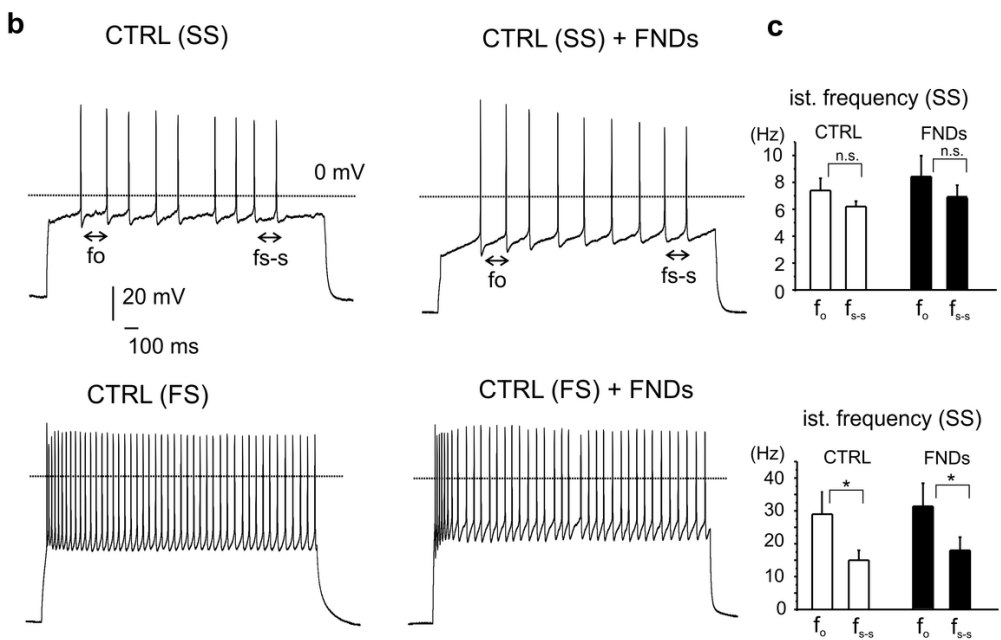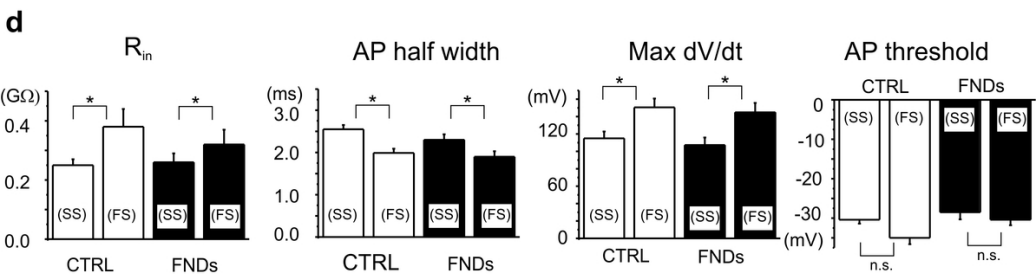

A

CTRL, without kink

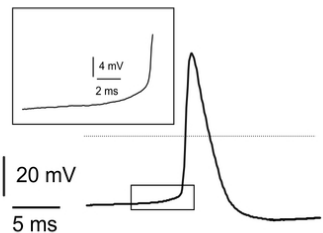
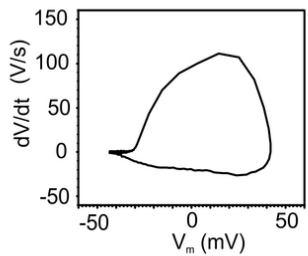

CTRL, with kink

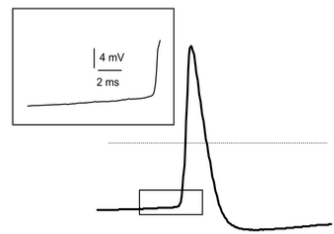
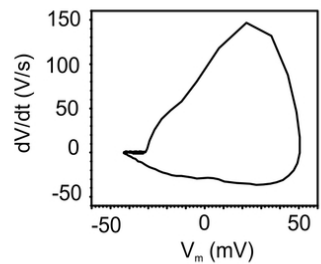

B

FNDs, without kink

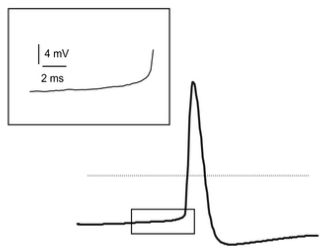
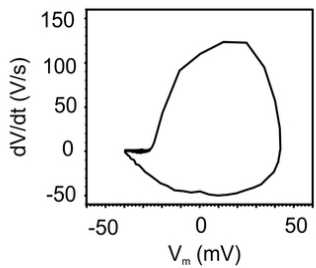

FNDs, with kink

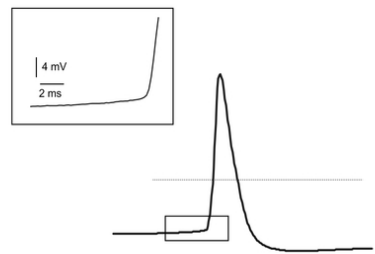
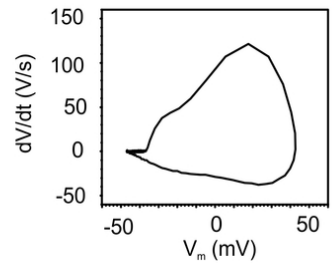

C

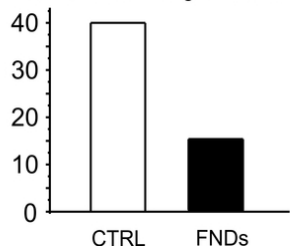

% cells exhibiting the KINK

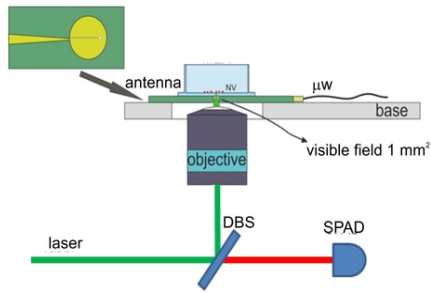

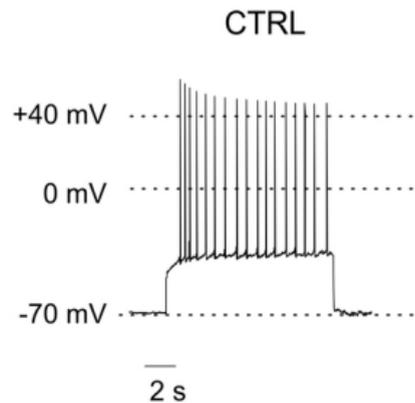

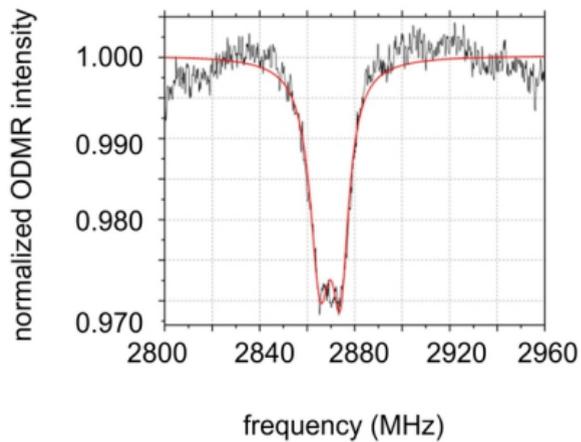

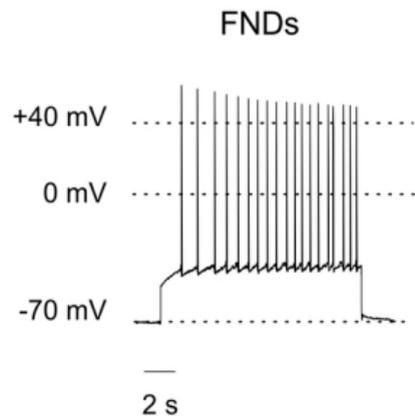